\documentclass[prb,twocolumn,floatfix]{revtex4}
\usepackage{graphics,epsfig,amsfonts,amssymb,amsmath,xcolor,ulem}
\begin{document}
\title{Interactions in the 8-orbital model for twisted bilayer graphene}
\author{M.J. Calder\'on}
\email{mariaj.calderon@csic.es}
\author{E. Bascones}
\email{leni.bascones@csic.es}
\affiliation{Instituto de Ciencia de Materiales de Madrid (ICMM). Consejo Superior de Investigaciones Cient\'ificas (CSIC), Sor Juana In\'es de la Cruz 3, 28049 Madrid (Spain).}
\date{\today}
\begin{abstract}  
{We calculate the interactions between the Wannier functions of the 8-orbital model for  twisted bilayer graphene (TBG).  In this model, two orbitals  per valley centered 
at the AA regions,  the AA-p orbitals, account for the most part of the spectral weight of the flats bands. Exchange and assisted-hopping terms between these orbitals are found to be small. Therefore, 
the low energy properties of TBG will be determined by the density-density interactions. These interactions decay with the distance much faster than in the two orbital model,
following a $1/r$ law in the absence of gates. The magnitude of the largest interaction in the model, the onsite term between the flat band orbitals, is controlled by the size of the AA regions 
and  is estimated  to be $\sim 40$ meV. To screen this interaction, the metallic gates have to be placed at a distance  $\lesssim 5$ nm. For larger distances only the long-range part of the interaction is substantially screened. The model reproduces the band deformation induced by doping found in other approaches within the Hartree approximation. Such  deformation reveals the presence of other orbitals
 in the flat bands and is sensitive to the inclusion of the interactions involving them.}
\end{abstract} 
\maketitle

\section{Introduction}
The observation of insulating and superconducting states in twisted bilayer graphene (TBG) and
other moir\'e systems have spurred the interest of the scientific community~\cite{CaoNat2018_1, CaoNat2018_2}. 
Moir\'e systems offer an unprecedented opportunity to control the phase diagram by tuning the twist angle, the
doping, the pressure, the screening by gates or the coupling to the 
substrate.~\cite{CaoNat2018_1,CaoNat2018_2,YankowitzScience2019,LuNat2019,SharpeScience2019,StepanovNat2020,SaitoNatPhys2020,LiuArXiv2020} 
On spite of the intense effort, the nature of these states is still unknown. A key underlying issue is whether the 
insulating states are Mott insulators or they can be described as    the result of a standard symmetry 
breaking state.~\cite{PizarroJphysComm2019,ChoiNatPhys2019,JiangNat2019,XieNat2019,hauleArXiv2019,CalderonNPJQMat2020}

Experimental evidence of different symmetry breaking states have been reported by several groups.~\cite{LuNat2019,SharpeScience2019,JiangNat2019,WongNat2020,ZondinerNat2020,CaoArXiv2020,SaitoArXiv2020} 
Theoretically, mean field calculations, mostly based on the continuum model but also on atomistic 
calculations,~\cite{XiePRL2020,BultinikArXiv2019,LiuArXiv2019,ZhangPRB2020,CeaPRB2020,GonzalezArXiv2020}  have found very small energy differences between distinct symmetry breaking phases. 
Even at the mean field Hartree level in the non ordered states, a strong doping dependence of the flat bands was 
predicted.\cite{RademakerPRB2018,GuineaPNAS2018,RademakerPRB2019,CeaPRB2019,GoodwinArXiv2020,NovelliArXiv2020} 
With electron doping, the Dirac points increase their energy with respect to the states at the center of the Brillouin 
zone $\Gamma$. The opposite change is found with hole doping. This effect is due to the charge inhomogeneity
as the doped electrons or holes accumulate at the center of the moir\'e unit cell, in the AA regions.

The emergence of the correlated states at the integer fillings of the moir\'e unit-cell  suggests that
the underlying physics could be easier to understand within interacting models based on a reduced
number of effective moir\'e orbitals. Such a formulation is specially important to disentangle the 
possible role of Mott physics in the correlated states, but it can be also very useful to address 
the symmetry breaking states.  Initial attempts to build a Wannier function based model 
 started from fittings which included only the flat bands, the so-called two orbital model.\cite{YuanPRB2018,KoshinoPRX2018,KangPRX2018} This model features
$p_+$ and $p_-$ orbitals centered at the vertices of the hexagonal  moir\'e unit cell,  the AB/BA domains.
$p_+$ and $p_-$ orbitals acquire  a phase $e^{\pm i 2 \pi/3}$ under a $C_3$ rotation. 
The resulting Wannier functions of this model had an unconventional spinner shape with the charge
distributed on three lobes at the AA regions, far from the wave-function center. As it was extensively 
discussed in the literature,\cite{PoPRX2018,ZouPRB2018,PoPRB2019,SongPRL2019,AngeliPRB2018} topological obstructions appear in  this model which prevent the
explicit inclusion of all the symmetries including the emergent ones, such as the charge conservation  in a given valley. 
Beyond the topological obstructions, the large overlap between Wannier functions centered at different 
sites results in interactions decaying very slowly with distance and in sizable exchange and 
assisted hopping terms which complicate the study of correlations,\cite{KoshinoPRX2018,KangPRX2018,GuineaPNAS2018,KangPRL2019} The assisted hopping terms were
required to explain the band deformation with doping found in the atomistic descriptions and the
continuum model at the Hartree level.~\cite{GuineaPNAS2018}

To avoid the topological obstructions, effective models including a larger number of orbitals per valley, from four 
to ten, have been proposed.~\cite{PoPRB2019,CarrPRR2019,SongPRL2019,CarrPRR2019-2} 
Beside the flat bands, these models include in the fittings other bands at higher energies above and/or 
below them. The orbital character of the flat bands is different at the Dirac points $K, K'$ 
and at $\Gamma$. It is $p_+$ and $p_-$ at the Dirac points and $s$ and $p_z$, respectively, at the two eigenvalues at $\Gamma$ . 
Here $s$ and $p_z$ transform trivially under a $C_3$ transformation and are defined with respect to the 
mirror symmetry $M_x$ which flips the $x$ coordinate producing a layer exchanging two-fold rotation in 3D. 
$p_z$ changes sign under  $M_x$ while $s$ does not~\cite{PoPRB2019}.
Among the models proposed, the eight band model is the only one which
is based on a Wannier function construction and, at the same time, avoids breaking heavily the approximate particle-hole 
symmetry including four bands below the Dirac points and four bands above. 
Unlike the spinner shape of the Wannier functions in the two orbital model, in this model the 
$p_+$ and $p_-$ orbitals are centered at the AA regions where the charge is concentrated. 
The Wannier functions and tight binding models were built to ensure the maximum weight of these
orbitals in the flat bands~\cite{CarrPRR2019-2}. The interactions between  these Wannier states, necessary
to address the correlation effects in TBG, are not available yet.

 Here we calculate the interactions between the Wannier functions of the 8-orbital model~\cite{CarrPRR2019-2} for TBG. We consider
 intra and inter-orbital density-density interactions in the same and different unit cells, as well as the exchange and
assisted hopping terms, and compute the screening of the interactions  by metallic gates. Special emphasis is made 
on the interactions involving the $p_+$ and $p_-$ orbitals (AA-p) centered at the AA region, as they will be instrumental in 
determining the correlated phases of TBG.  We find that  the density-density interactions decay with the distance 
much faster than in the two orbital model, following a $1/r$ decay 
in the absence of gates, and the exchange and assisted-hopping terms are small.   
The band deformation, found in other approaches\cite{GuineaPNAS2018,RademakerPRB2019,CeaPRB2019,GoodwinArXiv2020} within the Hartree approximation, 
is reproduced by this model and reveals the different orbital content of the flat bands at $\Gamma$ and $K$. Comparing the magnitude of this effect in our approach and
in atomistic calculations\cite{GoodwinArXiv2020}  
allow us to estimate the value of the dielectric constant for the eight orbital model   as $\epsilon \sim12$ for a free standing TBG. 
For this value of $\epsilon$, the largest interaction term in this model, the onsite Coulomb interaction between the AA-p orbitals $\rm U^0_{\rm AA-p,AA-p}\sim 42$ meV.
Nevertheless,  including the interactions involving the other orbitals is important for a proper description of the correlated states. 
$\rm U^0_{\rm AA-p,AA-p}$ is noticeably screened by metallic gates placed at $\sim 4-5$ nm. The primary 
effect of a gate placed at a larger distance is to screen the extended (long-range) part of the interaction.

In Sec.~\ref{sec:model} we summarise  the main characteristics of the model introduced in Ref.~[\onlinecite{CarrPRR2019-2}]. In Sec.~\ref{sec:interactions} the interactions among the 8 orbitals are calculated and the effect of the screening by nearby gates is shown in Sec.~\ref{sec:screening}. In Sec.~\ref{sec:hartree} the effect of the interactions on the bands at the Hartree level is discussed. Finally, we conclude in Sec.~\ref{sec:discussion}. Details of the calculations are given in the appendix.

\section{The model}
\label{sec:model}
\begin{figure}
\leavevmode
\includegraphics[clip,width=0.233\textwidth]{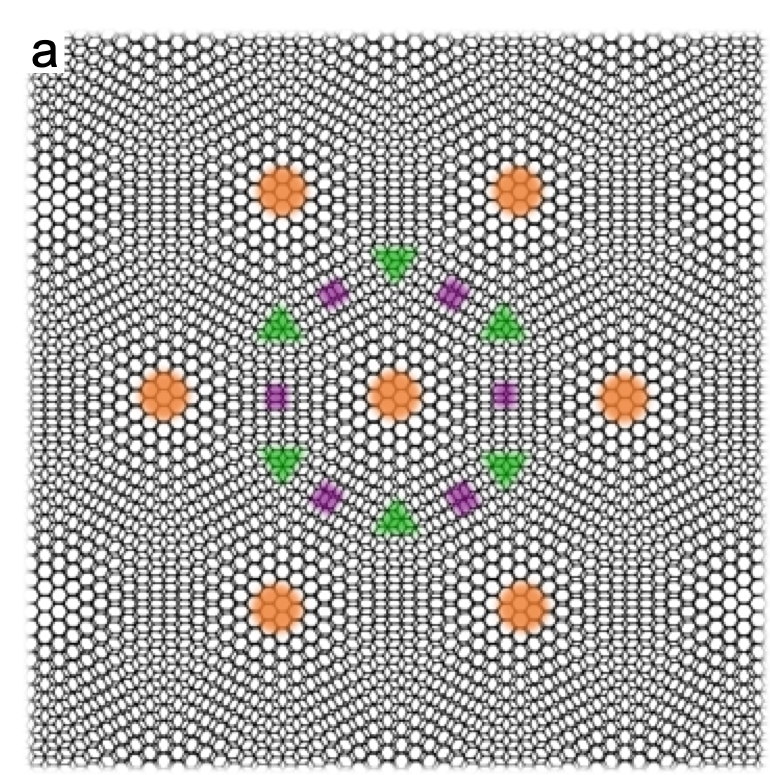} 
\includegraphics[clip,width=0.24\textwidth]{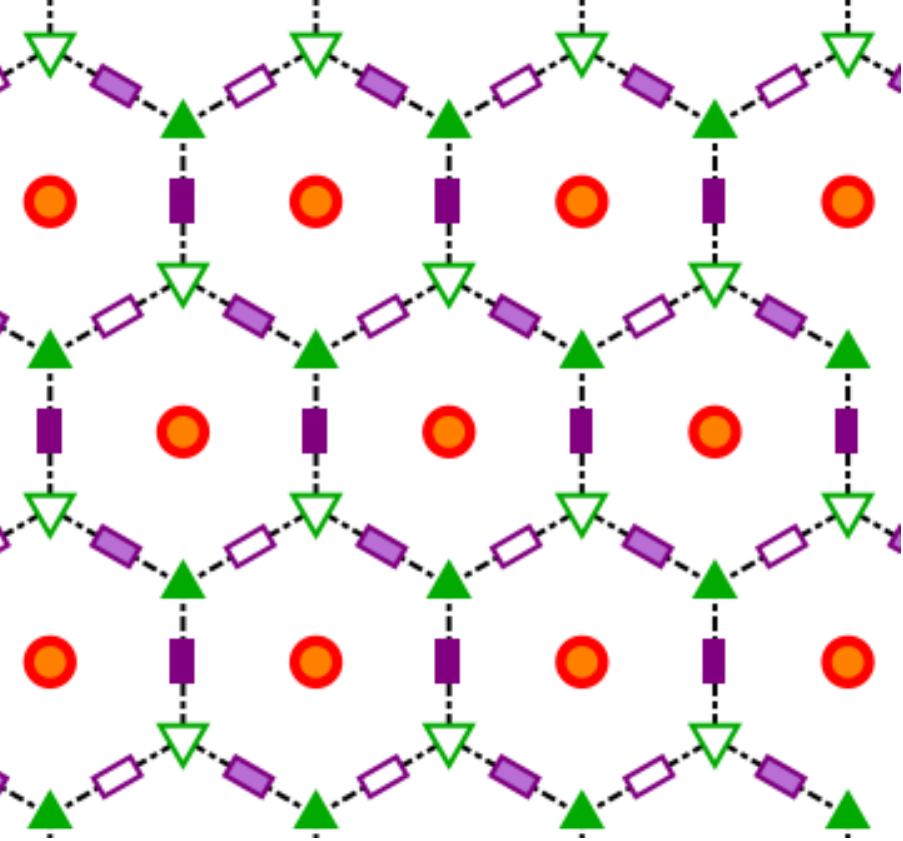} 
\includegraphics[clip,width=0.48\textwidth]{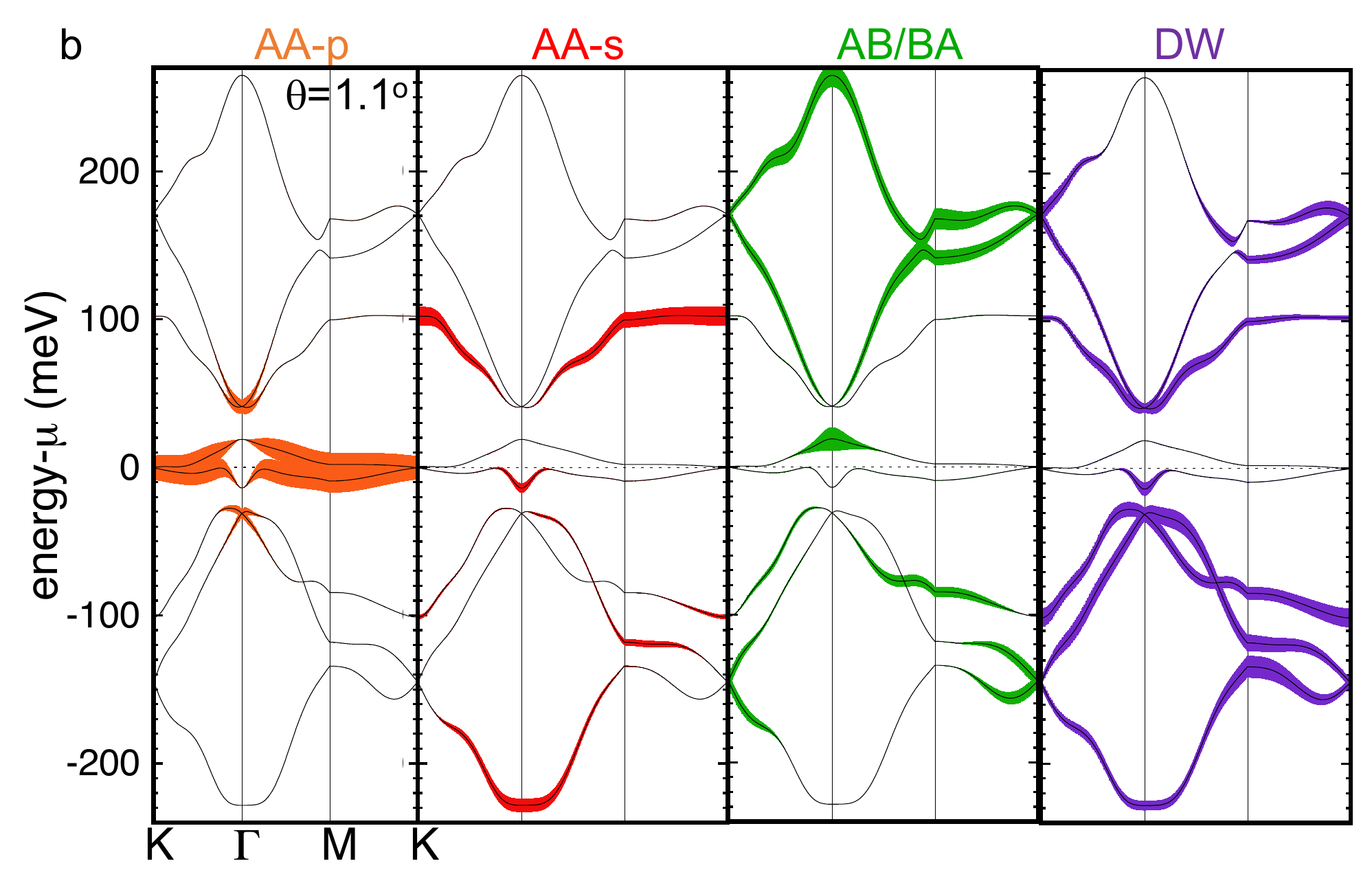}
\caption{(a)  Left: Top view of the stacking pattern in TBG with the AA, AB/BA and DW marked  following the color code on the right panel. Right: Effective moir\'e lattices where the Wannier functions are placed: Triangular AA-p (orange) and AA-s (red), hexagonal AB/BA (green) and Kagome DW (purple). Solid and empty symbols differentiate the inequivalent sites of the hexagonal and kagome lattices. (b) Non-interacting bands for $\theta=1.10^{\rm o}$. The color width in each panel indicates the corresponding orbital weight for each band.}  
\label{fig:lattice} 
\end{figure}

We consider the 8-band model per valley for TBG calculated in Ref.~[\onlinecite{CarrPRR2019-2,kaxiraskp,kaxiras-github}].
Taking into account the valley degree of freedom, the model includes 16 spin degenerate orbitals. 
The model is free from topological obstructions:
all the symmetries of the TBG are incorporated explicitly in the model, including the emergent ones.
The Wannier functions and tight-binding parameters are defined for a single valley. The ones corresponding to the other valley
are obtained by time-reversal symmetry. Within each valley the model includes the flat bands  as well as three bands above and three bands below them,  named here as higher energy bands, see Fig.~\ref{fig:lattice}. 
The orbitals in the two valleys are not coupled by kinetic energy, but they are  coupled by interactions. 

\begin{figure*}
\leavevmode
\includegraphics[clip,width=0.32\textwidth]{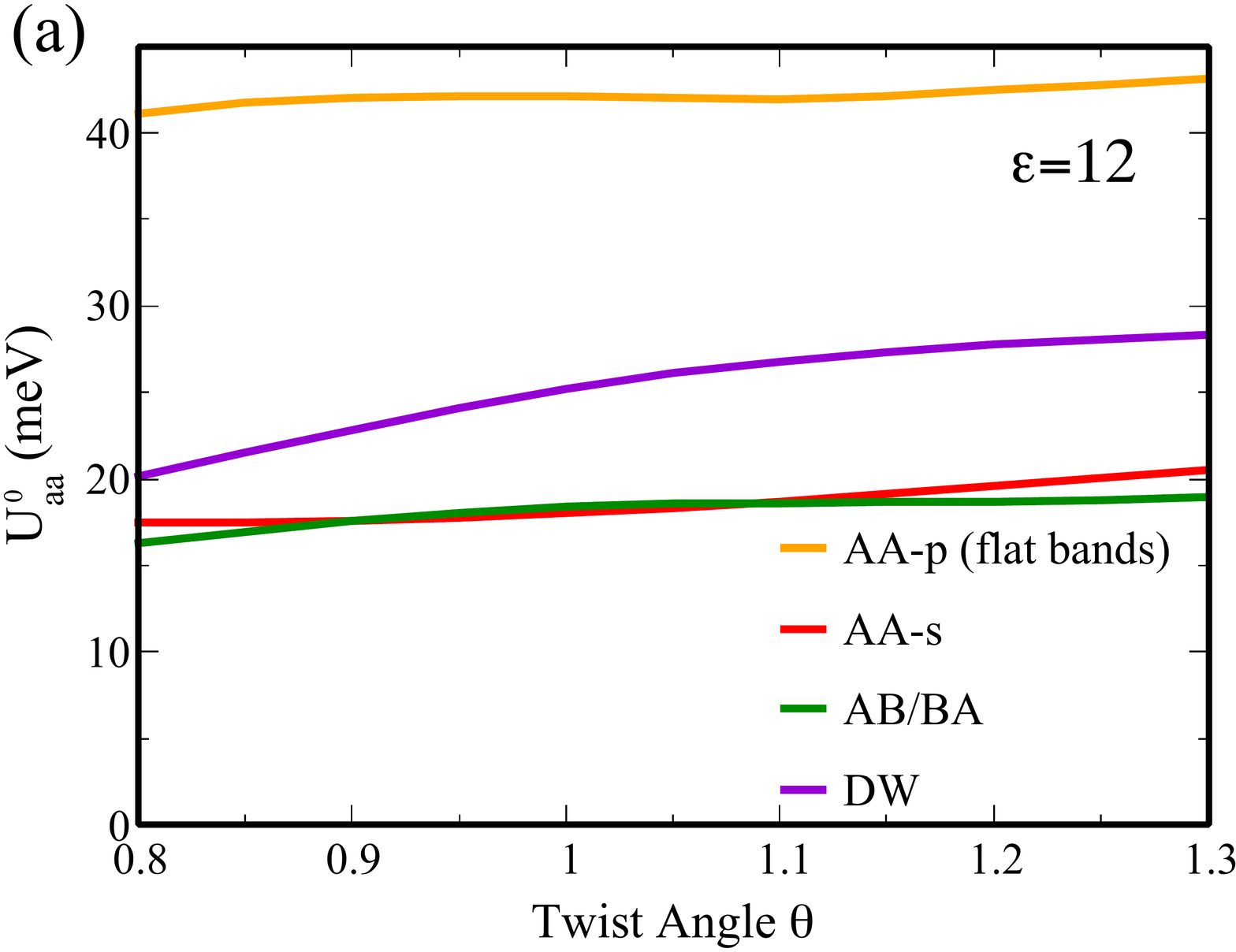}
\includegraphics[clip,width=0.32\textwidth]{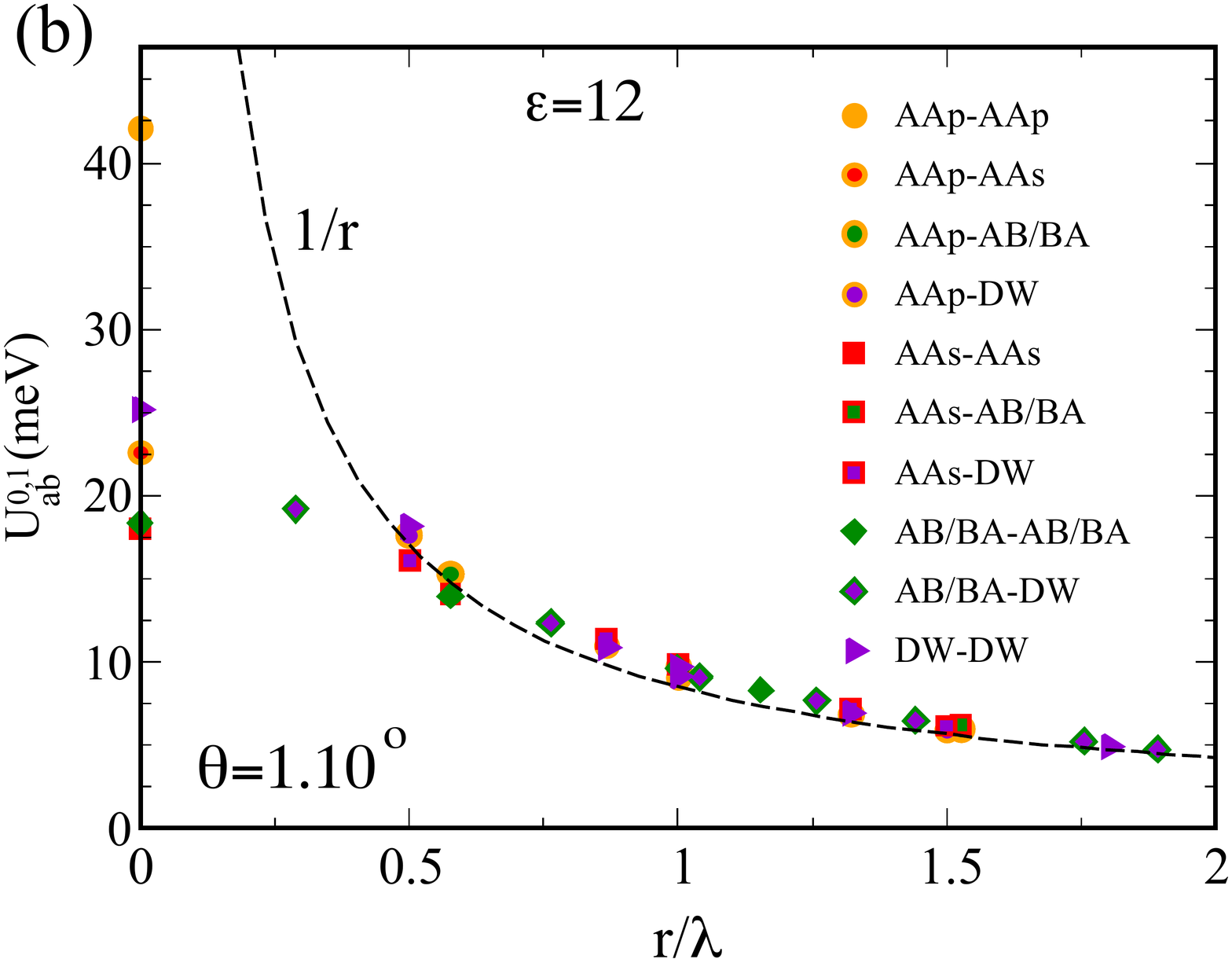}
\includegraphics[clip,width=0.32\textwidth]{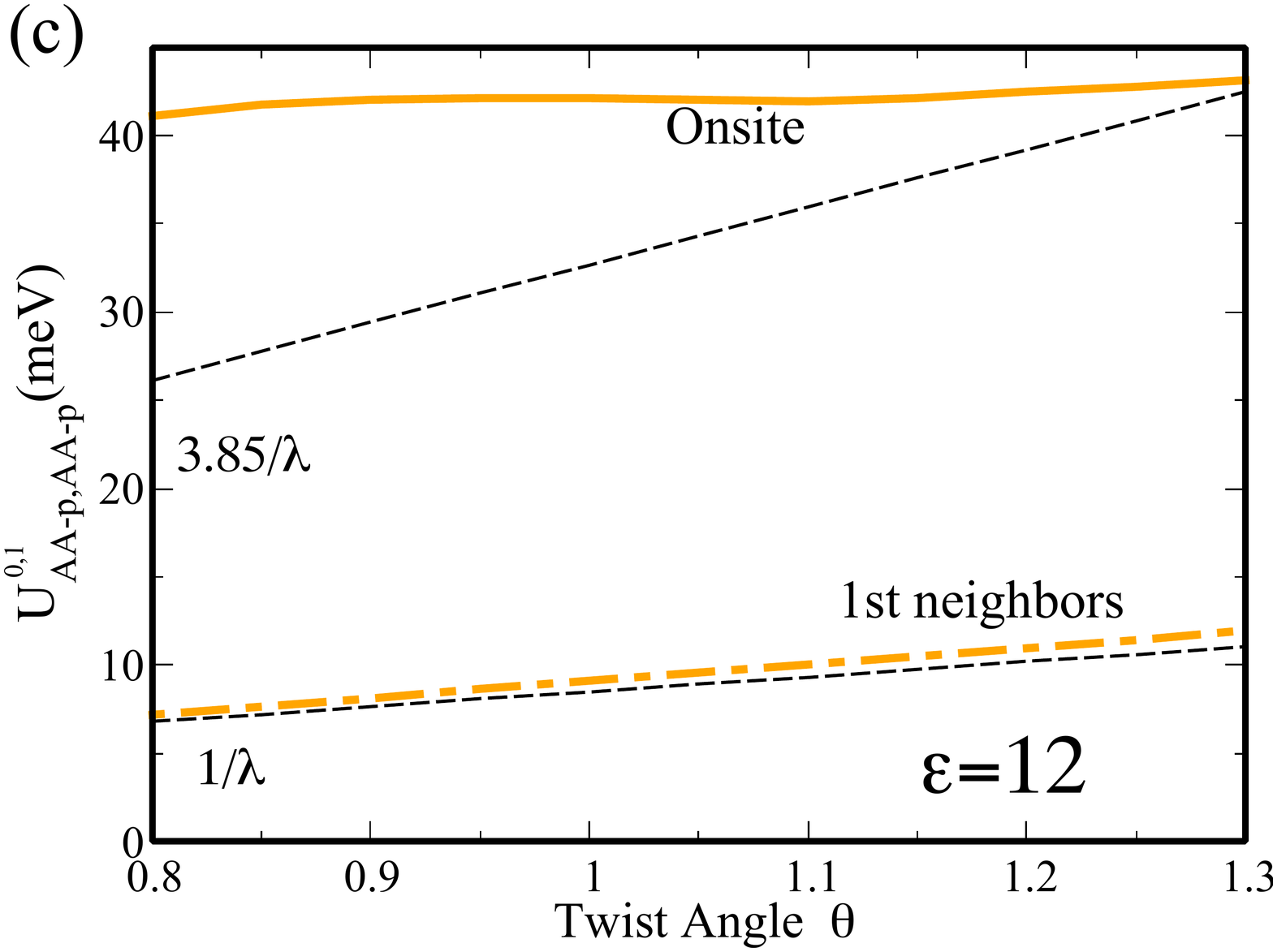}
\caption{(a) Intraorbital  onsite interaction for the four kinds of orbitals as a function of the twist angle $\theta$, assuming $\epsilon=12$. (b) Interactions arising between the eight orbitals as a function of distance for $\theta=1.1^{\rm o}$. $\lambda$ is the moir\'e lattice constant. For distances $r>\lambda/2$, the interactions fit the $1/r$ decay of the unscreened potential. (c) Comparison of the onsite and 1$^{\rm st}$ neighbors interaction for the AA-p orbitals as a function of twist angle.}  
\label{fig:interactions} 
\end{figure*}

The eight orbitals in each valley are centered at the three symmetry lattices defined in the TBG, see Fig.~\ref{fig:lattice}, with the location and the type of orbital
 determined by symmetry considerations, see Refs.~[\onlinecite{PoPRB2019,CarrPRR2019-2}].  Orbitals 1 to 3 are centered at the triangular
lattice, formed by the AA regions. Orbitals 1 and 2, named AA-p, have $p_+$ and $p_-$ symmetry, and orbital 3, AA-s, has $s$ symmetry. Orbitals 4 and 5 
have $p_z$ symmetry and are centered at the hexagonal lattice formed by the AB/BA domains.  They are named hexagonal AB/BA 
orbitals. Finally, orbitals 6, 7 and 8 (DW) have $s$-symmetry and form a Kagome lattice, located at the center of the domain walls  which separate the AB/BA domains and join two AA regions. 
In all figures we assign the color orange to AA-p, red to AA-s, green to AB/BA and purple to DW orbitals. 

The Wannier functions used here have been derived from fits to the  low energy band structure of a  realistic generalized $k.p$ model for relaxed TBG ~\cite{CarrPRR2019,CarrPRR2019-2,kaxiraskp}   
and they provide good fitting for twist angles $\theta$ between $0.6^{\rm o}$ and $1.3^{\rm o}$ but similar 
Wannier functions could be obtained for non-relaxed models. 
The relaxation reduces the size of the AA regions, increasing the AB and BA domains area, and produces vertical corrugation.~\cite{UchidaPRB2014,NguyenPRB2017,AngeliPRB2018,ZhangJMPS2018,CarrPRR2019,yooNatMat2019}

As shown in Fig.~\ref{fig:lattice} more than 90$\%$ of the spectral weight of the flat bands, including the Dirac points, have AA-p ($p_+$ and $p_-$) character. 
Therefore, the AA-p orbitals will be key players in the correlated states of TBG. 
On the other hand, the lower (upper) flat band at $\Gamma$ has $s$ ($p_z$) character above the twist angle $ \theta_0\sim 1^{\rm o}$ at which the two flat bands touch.~\cite{CarrPRR2019} 
 Below this angle, the orbital character of the two flat bands at $\Gamma$ is exchanged.
For more details on the tight-binding model and on the Wannier functions we refer the reader to Ref.~[\onlinecite{CarrPRR2019-2,kaxiraskp,kaxiras-github}]. 

\section{Interactions}
\label{sec:interactions}
We calculate the interactions between the electrons in the Wannier functions belonging to any of the two valleys. Details on the procedure are given in the Appendix. Except otherwise indicated, in order to compute the interactions between electrons in different orbitals we assume an interaction of the form $V(r)\sim e^2/(4 \pi \epsilon_0 \epsilon r)$. Here  $\epsilon$ is an effective dielectric constant for the 8-band model and $\epsilon_0$ the permittivity of vacuum. The value of the interactions is proportional to the value of $1/\epsilon$ used. In the figures we use $\epsilon=12$ as estimated from the comparison with atomistic calculations for free standing TBG in  Section~\ref{sec:hartree}.  $\epsilon$ is expected to be larger if the TBG is encapsulated by boron nitride (h-BN)\cite{GoodwinPRB2020} $\epsilon \sim \epsilon_{TBG}+\epsilon_{h-BN}-1 \sim 15$ leading in such a case to a reduction of the interactions of approximately a 20$\%$.

\subsection*{Density-density interactions}

Fig.~\ref{fig:interactions} (a) shows  $\rm U^0_{aa}$, the interactions between two electrons located  in the same orbital and site (intraorbital and onsite),  as a function of the twist angle $\theta$. Here $a$ refers  to the kind of orbital. 
$\rm U^0_{\rm AA-p,AA-p}$,   doubles approximately the interaction of the other orbitals, consistent with their different spreads.~\cite{CarrPRR2019-2} 
Nevertheless, not only the radii of the orbitals determine the interaction $\rm U^0_{a,a}$ but also their shape matters. For instance, $\rm U^0_{\rm AA-s,AA-s}$ is smaller than 
the DW interaction, on spite of the larger radius of the latter.~\cite{CarrPRR2019-2} The larger value of $\rm U^0_{AA-p,AA-p}$ emphasizes the predominant role played by these orbitals in the correlation effects in 
TBG, but the  AA-s, AB/BA and DW onsite interactions are sizable $\sim 20-30$ meV.

\begin{figure}
\leavevmode
\includegraphics[clip,width=0.43\textwidth]{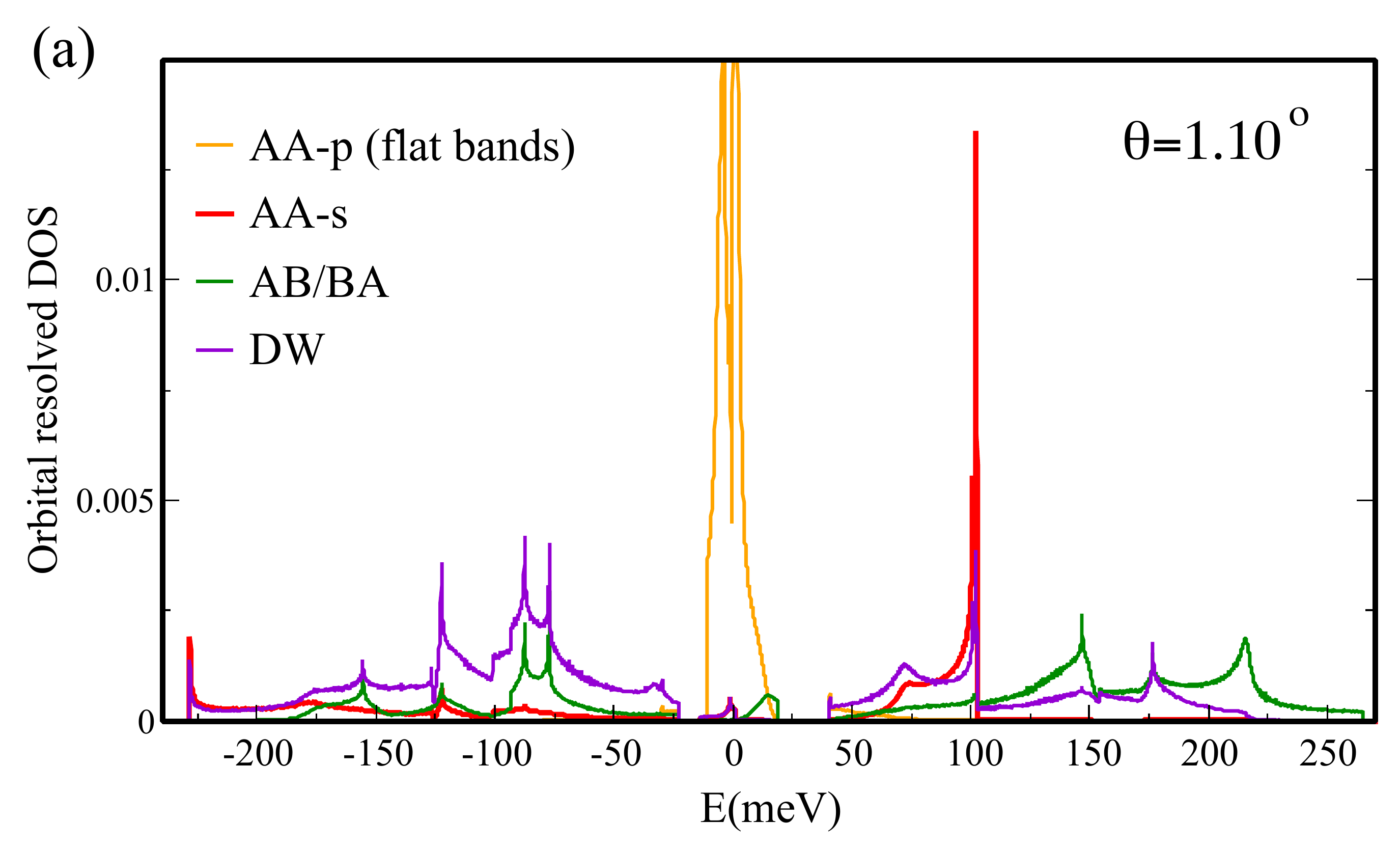}
\includegraphics[clip,width=0.43\textwidth]{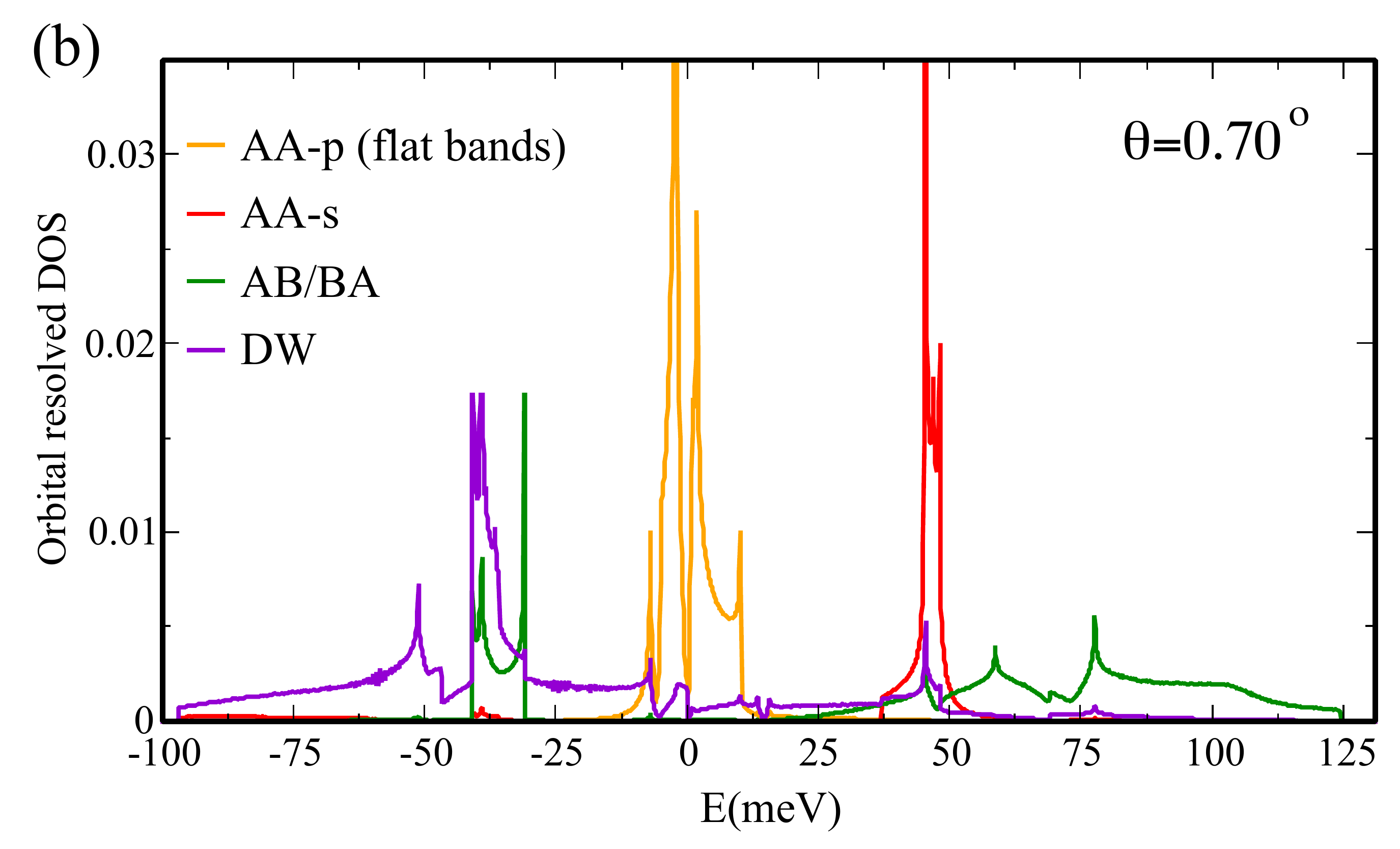}
\caption{Non-interacting orbital resolved density of states for each kind of orbital for (a) $\theta=1.10^{\rm o}$ and (b) $\theta=0.70^{\rm o}$.}  
\label{fig:dos} 
\end{figure}

 The correlated character of an orbital depends not only on the magnitude of the interactions involved but also on its bandwidth. The non-interacting orbital resolved density of states (DOS) 
for $\theta\sim 1.10^{\rm o}$ is shown  in Fig.~\ref{fig:dos} (a), with the DOS corresponding to the same kind of orbitals, for instance, the three DW orbitals, added. 
The AA-p orbitals present the narrowest band with their bandwidth $\rm W_{AA-p} \sim 20-30$ meV, smaller than their onsite interaction energy $\rm U^0_{AA-p,AA-p}$. Therefore, Mott physics 
can play a significant role in their electronic spectral weight.  The larger bandwidths $\sim 200-400$ meV of the triangular AA-s, hexagonal~AB/BA and 
Kagome~DW orbitals suggest that Mott localization is unlikely for them at this twist angle $\theta\sim 1.10^{\rm o}$. Nevertheless, peaks in the density of states of these 
orbitals are prominent, particularly, for the triangular AA-s orbital. Such peaks could result in correlated states driven by Fermi surface instabilities.

Interesting changes appear in the orbital resolved DOS when the twist angle is reduced, as observed in Fig.~\ref{fig:dos} (b) for $\theta\sim 0.70^{\rm o}$ with a generic bandwidth 
reduction and higher peaks in the DOS. These changes correlate well with the evolution of the bands, which are narrower and show flat bands   away from the 
charge neutrality point,~\cite{CarrPRR2019-2} and are accompanied by a noticeable spectral weight redistribution.  For an undoped $\theta\sim 1.10^{\rm o}$,  around 30$\%$ of the triangular AA-s orbital 
is below the Fermi level  and is spread over $\sim$150 meV while the rest of the spectral weight is concentrated in a range of energies of only 40~meV. 
On the other hand, for $\theta\sim 0.70^{\rm o}$ most of the spectral weight of the AA-s orbital has been transferred to the unoccupied bands which concentrate up to 90$\%$ of this weight in 
only $\sim$20 meV, such that the ratio between the onsite interaction and the bandwidth becomes favorable for Mott correlations away from the two flat bands at the CNP.   Notably, insulating states
 have been found at high integer dopings well below the magic angle.~\cite{CodecidoSA2019} 

In addition to the intraorbital onsite interactions $\rm U^0_{a,a}$, electrons in a given orbital are expected to interact with the other orbitals centered in the same or different unit cells. 
Interactions between the orbitals   in the same and neighboring cells are plotted in Fig.~\ref{fig:interactions} (b) for $\theta=1.10^{\rm o}$ as a function of their distance in units of $\lambda$, the  moir\'e lattice constant. For distances larger than $\lambda/2$ their magnitudes 
approximate the expectations of an $1/\epsilon r$ dependence for point-like charges. 
This is the case of  the interactions between orbitals centered at the same site (at zero distance) such as the U$^0_{a,a}$ discussed above, the interactions between two different AA-p 
orbitals or  those between an AA-p orbital and the AA-s orbital $\rm U^0_{AA-p,AA-s}$, but also of  the interaction between the closest hexagonal and kagome orbitals at 
$r=\lambda/(2\sqrt{3})$.  To our level of 
accuracy, the interactions between any two AA-p orbitals are  independent of whether the two orbitals involved have the same or different valley and  $p_+$ or $p_-$ character. The onsite interaction between 
electrons in AA-s and those in AA-p orbitals is larger than the intraorbital $\rm U^0_{AA-s,AA-s}$. 

AA-p orbitals are expected to play a key role in the correlated states of TBG. Fig.~\ref{fig:interactions}(c) compares the dependence on the twist angle of the interactions between these orbitals 
when they are in the same or in different unit cells. In the range of angles plotted the interaction in neighboring unit cells ${\rm U^1_{\rm AA-p,AA-p}}$ decays as $1/\lambda$ while
the onsite interaction ${\rm U^0_{ AA-p,AA-p}}$ is approximately constant. ${\rm U^0_{\rm AA-p,AA-p}}$ does not scale with the size of the moir\'e unit cell, but with the size of the AA region, which tends to  a constant value below twist angles 
 $\theta \sim 1^{\circ}-2^{\circ}$~[\onlinecite{NguyenPRB2017,AngeliPRB2018,ZhangJMPS2018,yooNatMat2019}]. Approximating  
 ${\rm U^0_{AA-p,AA-p}} \sim e^2/(4 \pi \epsilon_0 \epsilon_r {\rm R_{AA-p}})$  gives a characteristic length scale 
 ${\rm R_{AA-p} \sim 28} $ {\rm {\AA}} for ${\rm U^0_{AA-p,AA-p}}$ 
 close to the effective radius of the AA region $\sim 40$ {\rm {\AA}} estimated in other calculations.~\cite{ZhangJMPS2018} Decreasing the twist angle makes the interaction between the
electrons in AA-p orbitals less extended.

\subsection*{Exchange and assisted hopping interactions} 
 Exchange interactions J and assisted hopping terms are important in the two orbital model of TBG.~\cite{KoshinoPRX2018,GuineaPNAS2018, KangPRX2018} In the model considered here, we find these interactions to be much smaller than the density-density ones. In particular, this is the case for all the terms involving the AA-p orbitals, predominant in the flat bands. 
Different reasons account for the small values of the exchange interactions. In the onsite cases in which the envelope functions have a large overlap, there is destructive interference damping 
the interaction. For instance,  the Hund's coupling between electrons in different valleys involve a product of Bloch factors oscillating at the atomic scale which integrate to a negligible value ($\sim 0.01$ meV), 
while the AA-p intravalley Hund's coupling is dramatically reduced ($\sim 0.05$ meV)  as the product of the envelope functions changes sign within the unit cell. 
The small value of the exchange and assisted hopping terms between AA-p orbitals in different unit cells ($\sim 0.1$ meV and $\sim 0.2$ meV) is due to the weak 
overlap between the orbitals.

The small exchange and assisted hopping interactions between the flat band orbitals in the 8-orbital model contrast with the findings in the two orbital 
model, where these terms were sizable.  The differences between both models originate in the large overlap between the spinner functions centered at different sites in the two orbital model.

 Larger values are found for the exchange and assisted hopping terms involving not only the AA-p but also other orbitals, for instance, the exchange interaction 
$\rm J^0_{\rm AA-p, AA-s} \sim $ 3 meV.  However, the contribution of AA-s to the flat bands is small and its spectral weight is spread over more than 200 meV. Therefore, it is unlikely that these exchange terms play an 
important role in the correlated states of TBG. The assisted hopping terms between the AA-p orbitals and the AB/BA and DW orbitals at the vertices and edges of the unit cell can reach a similar magnitude (2-3 meV). Such terms, though still small compared with the density-density interactions,  could induce a certain doping dependence of the band structure.

\section {Screening}
\label{sec:screening}
Recent experiments have shown the possibility to tune the phase diagram by changing the distance between the TBG and a metallic gate or controlling 
in-situ the doping of a nearby graphene bilayer.~\cite{StepanovNat2020,SaitoNatPhys2020,LiuArXiv2020} The screening reduces the interaction between the electrons and 
the effects of correlation are expected to weaken.~\cite{PizarroPRB2020,GoodwinPRB2020,StepanovNat2020}  The moir\'e lattice constant $\lambda$ is usually assumed to 
be the gate distance below which the interactions are considerably reduced.

\begin{figure}
\leavevmode
\includegraphics[clip,width=0.45\textwidth]{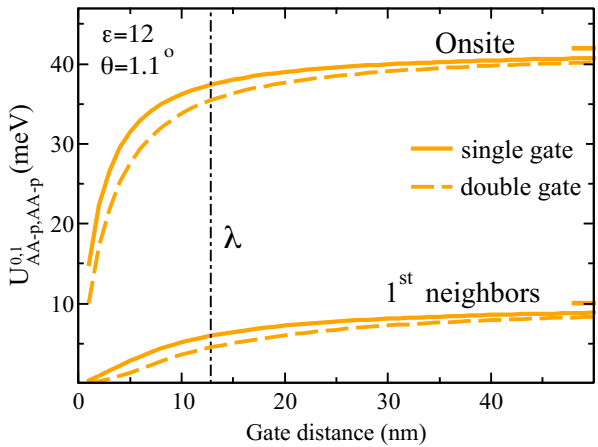}
\caption{
Interaction between AA-p orbitals in the same and in neighboring unit cells as a function of the distance to a single and double metallic gate for $\theta=1.1^{\circ}$ 
and $\epsilon=12$.  The small marks at the right vertical axis indicate the value of the interactions in the absence of metallic gates and the black dashed-dot vertical line signals the moir\'e lattice constant 
$\lambda$. 
}  
\label{fig:gates} 
\end{figure}

Fig.~\ref{fig:gates} shows the variation of the intra-orbital onsite and first nearest neighbor interactions, $\rm U^0_{AA-p,AA-p}$ and $\rm U^1_{AA-p,AA-p}$ as a function of the distance $d$ to a single or a double metallic gate
for a twist angle $\theta = 1.10^{\rm o}$. 
 The expressions for the interaction potential used are given in the Appendix.
When the distance to the gates is the moir\'e lattice constant  $d=\lambda (\theta = 1.10^{\rm o})=12.8$ nm the onsite interaction is reduced by
only an $11\%$ ($15\%$) for the case of a single (double) gate with respect to its non-screened value. It is necessary to place the gate at distances below 4 nm (single gate) or 6 nm (double gate) in order to reduce $\rm U^0_{AA-p,AA-p}$ by a 30\%.
The inter-cell interaction is, on the other hand, reduced, by a $41\%$ and $55\%$ respectively, when single and double gates are located at a distance $d=\lambda$. This indicates that the main
effect of a gate at a distance $d \sim \lambda$ is to reduce the range of the interaction,  while the onsite value is mainly unaffected.

At first sight our results could seem contradictory with the ones in the two orbital model, which showed a reduction by a factor of two of the onsite interaction for $d=10$ nm.~\cite{GoodwinPRB2020,StepanovNat2020}
Again,  the differences can be traced back to the Wannier functions. In the 8 orbital model most of the charge of the AA-p orbitals is contained in a single AA region. On the other hand in the two-orbital model, the spinner function has its charge spread over three AA-regions and the onsite interaction has contributions from the interaction between charges in the same and in neighboring AA regions.  In the absence of a metallic gate the interaction between the charges in different AA regions gives approximately half of the onsite interaction. It is this contribution, and not the intra-cell one,  which is primarily screened by a metallic gate placed at 10 nm.~\cite{GoodwinPRB2020}

\section{Hartree correction}
\label{sec:hartree}

As seen above, the correlation effects in TBG are dominated by the density-density interactions. Taking only the Hartree contribution of these interactions constitutes 
the simplest approximation in which the correlation effects can be introduced. The interacting part of the Hamiltonian enters as an orbital dependent
 shift of the onsite potential $V_{{\rm H},a}$ which depends on the interaction between the orbitals and their occupation. 
\begin{equation}
{\rm H_{Hartree}}=\sum_{a,\sigma} V_{{\rm H},a} n_{a,\sigma} 
\label{Eq:hartree1}
\end{equation}
with $n_a$ labelling the density of orbital $a$. The sum in $a$ includes the orbitals in both valleys and 
\begin{equation}
V_{{\rm H},a}=\frac{1}{2}\sum_{a,b,m_j}{\rm U_{ab}^{j}}\left( \langle n_b \rangle - \bar n \right)
\label{Eq:hartreepot}
\end{equation}
with the orbital dependent density $ \langle n_b \rangle$ calculated self-consistently. The sum is running to all the orbitals $b$ in the neighboring cells $m_j$  located at a distance $j$ times $\lambda$. 
In the summation above we exclude the contribution from orbitals $b=a$ centered at the same unit cell with the same spin, see the Appendix, 
and we have assumed that the spin and the translation symmetries at the moir\'e length scale are not broken.
We ensure charge neutrality over the unit cell to avoid double counting an effect which is already  included in the
tight-binding bands. We deduct from the density  $\bar n=\frac{1}{2}+\frac{\nu}{32}$ with $\nu$ the electrons (holes) doped to the TBG with respect to the charge neutrality point (CNP).  That is, we subtract the 
contribution produced by a uniform filling of all the orbitals, the closest approximation in our model to a uniform charge density through the moir\'e unit cell.

\begin{figure}
\leavevmode
\includegraphics[clip,width=0.48\textwidth]{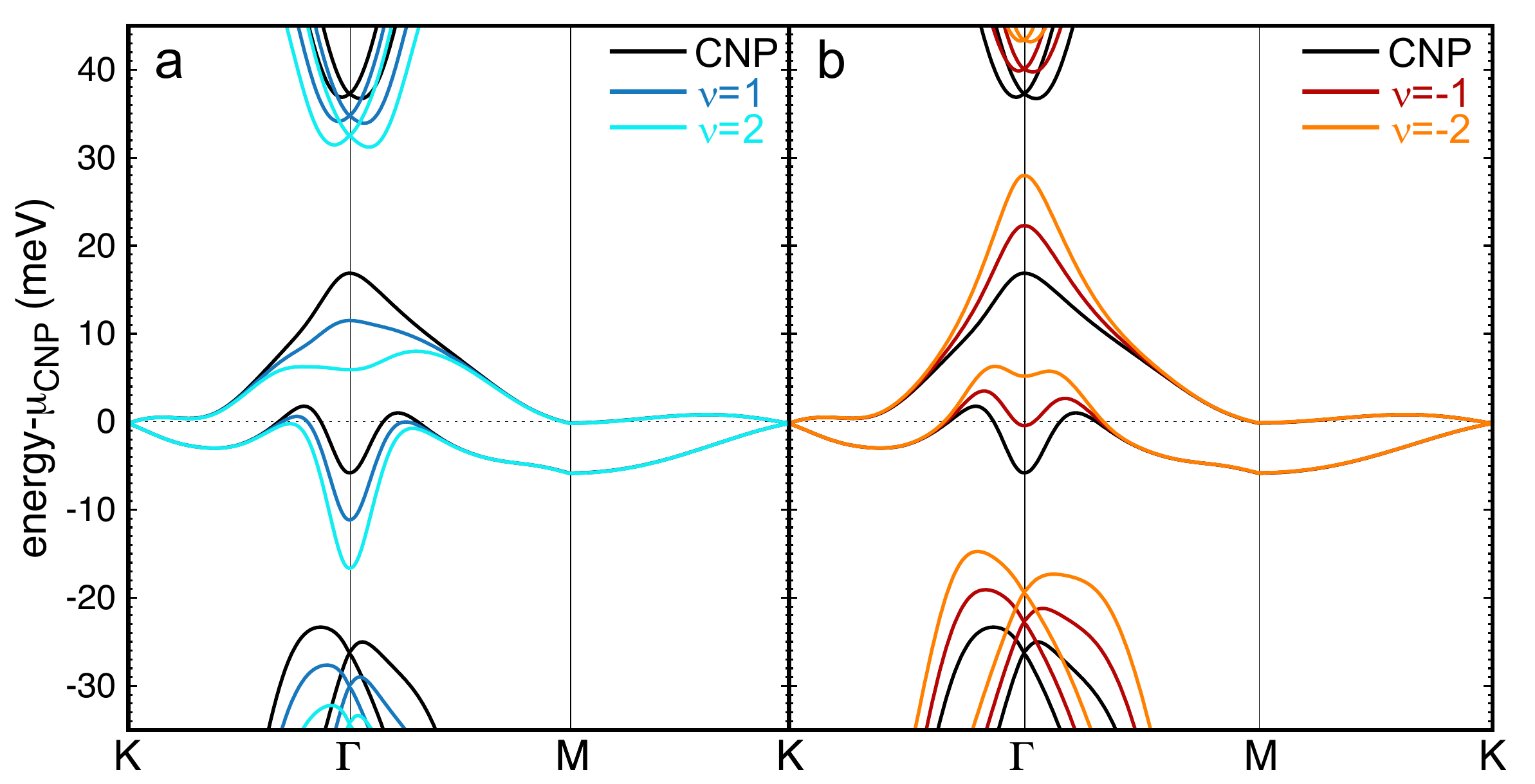}
\includegraphics[clip,width=0.245\textwidth]{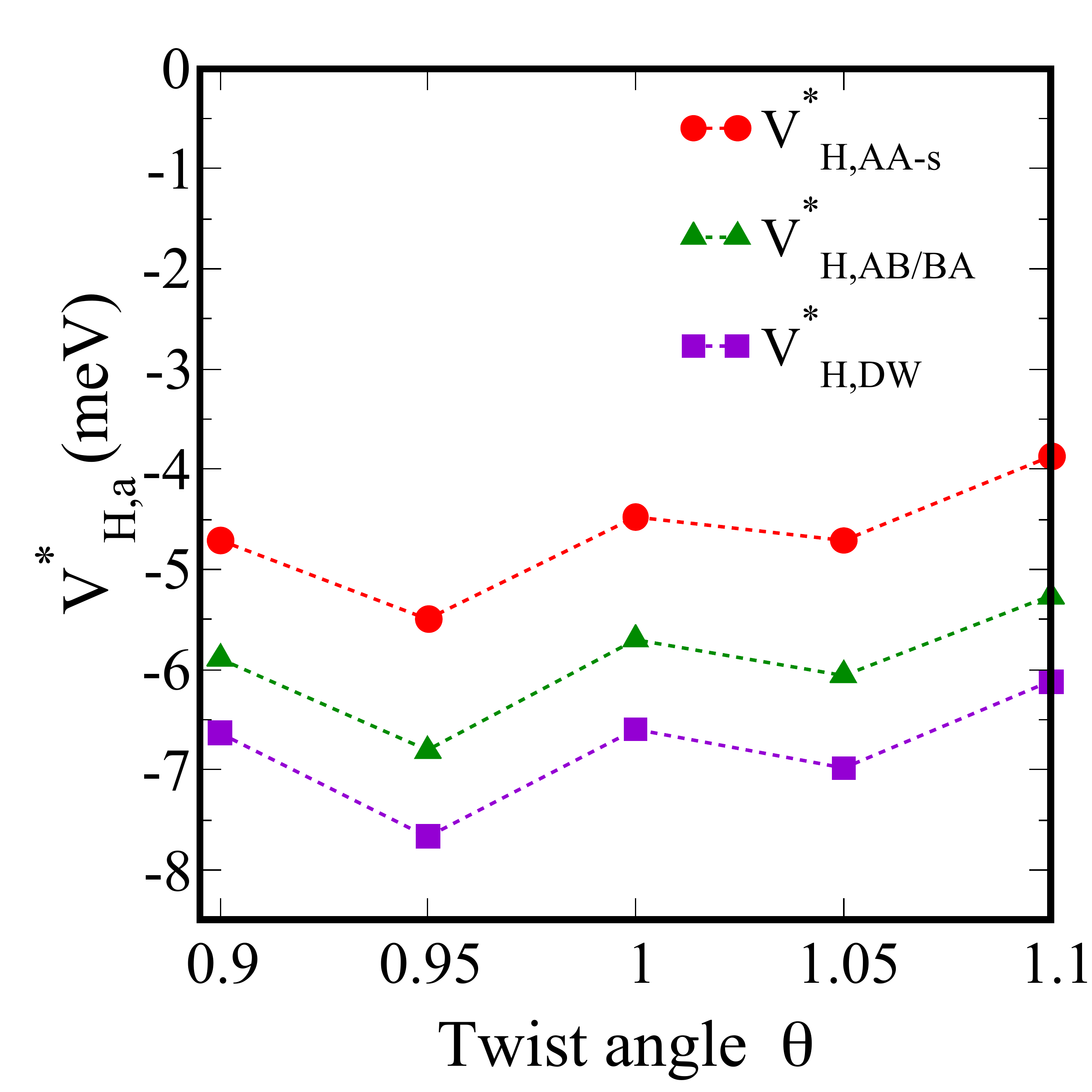}
\hskip -0.3 cm
\includegraphics[clip,width=0.245\textwidth]{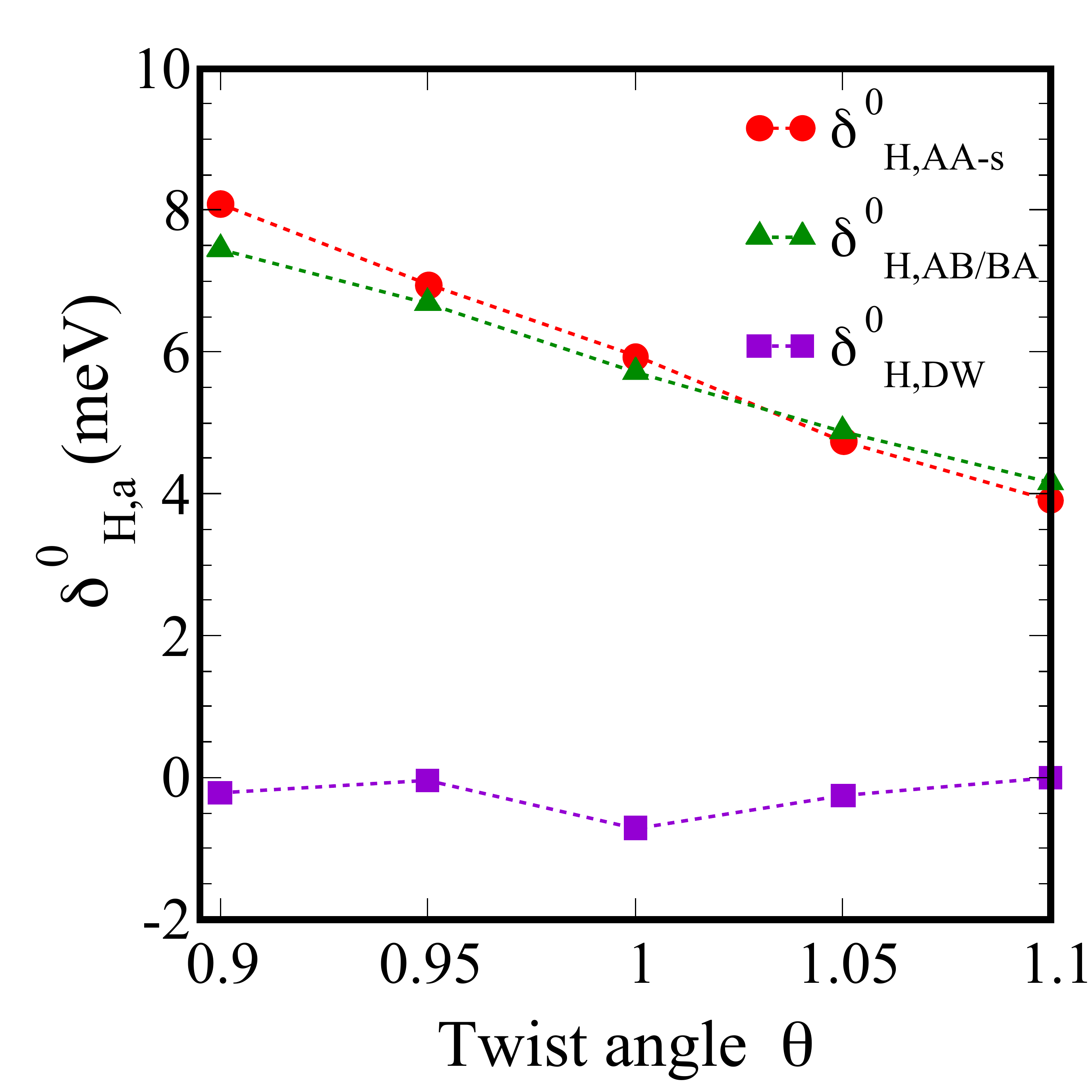}
\caption{Electronic bands in the Hartree approximation  corresponding to (a) an electron doped and (b) a hole doped  TBG with $\theta=1.05^{\circ}$. The energy of the Dirac points at each doping is used as a reference to facilitate the comparison. The Hartree correction has been taken into account in the bands for all the dopings, including the CNP.   (c) and (d) Parameters obtained from the fittings to the Hartree correction using Eq. (\ref{eq:hartreefit}), see text.  $\epsilon=12$ is used in the figure.}  
\label{fig:hartree} 
\end{figure}

In many materials, the Hartree potentials $V_{\rm {H,a}}$ vanish because  the last two terms cancel or can be included in a redefinition of the chemical potential of the effective model. 
This is not possible in TBG where the non-uniformity of the charge doped produces a doping dependent band 
deformation.~\cite{RademakerPRB2018,GuineaPNAS2018,RademakerPRB2019,CeaPRB2019,GoodwinArXiv2020}
Our interacting 8-band model reproduces this band deformation. Fig.~\ref{fig:hartree} (a) and (b) show the changes in the band shape  obtained in a self-consistent calculation when a TBG with angle $\theta= 1.05^{\circ}$ is doped.  
As in the other approaches~\cite{RademakerPRB2018,GuineaPNAS2018,RademakerPRB2019,CeaPRB2019,GoodwinArXiv2020} the changes in the band shape are opposite for electron and hole doping. When electrons (holes) are doped into the system the electronic states at $\Gamma$ decrease (increase) their energy with respect to the Dirac points and to the states at M. 

Contrary to findings in the two-orbital model, in the present 8-band model it is not necessary to introduce assisted hopping terms to reproduce these effects. They are associated to more conventional density-density interactions and reveal the different orbital contribution at 
$\Gamma$ and K, as early proposed in Ref.~[\onlinecite{RademakerPRB2018}].  The relative shift between the states in the flat bands at $\Gamma$ and those at M 
and K can be understood in terms of the different potential felt by each orbital. With doping, the electrons added  or removed are concentrated 
primarily at the AA regions in the AA-p orbitals  and the Hartree potential felt by these orbitals $V_{\rm{H,AA-p}}$ increases (decreases) with respect to the one felt by the other orbitals. Consequently the regions 
in k-space where the bands have  larger AA-p orbital weight increase (decrease) their energy. 

A similar shift, and with the same origin, of the states at  $\Gamma$ of the higher energy bands with respect to the Dirac points can be also appreciated in  Fig.~\ref{fig:hartree}. At large dopings and small $\epsilon$ the gap between the flat bands and these higher energy bands, whose $k$-dependence is also affected, could even close. Much smaller, but still finite, is the change in the flatband bandwidth at $\Gamma$.
The latter is due to the differences in the Hartree potencial of the hexagonal $p_z$  AB/BA and the ones of the $s$ orbitals AA-s and DW and may shift slightly the angle $\theta_0$ at which the orbital character of the two eigenvalues at $\Gamma$ is exchanged.

The band distortion is accompanied by a small charge redistribution with respect to the tight binding predictions. For instance, in the absence of interactions, in a TBG with $\theta=1.10^\circ$ doped with 3 electrons the 96\% of the extra charge would have been added to the AA-p orbitals. After self-consistency, only 93\% of the added electrons fills these orbitals. The difference has gone primarily to the DW orbitals. The redistribution is   slightly larger for a $\theta=0.95^\circ$ TBG with smaller bandwidth. For the same doping, 93\% of the doped charge is expected to populate the AA-p orbitals, but only 85\% is added to them after self-consistency.

For practical purposes, in our model it is convenient to define the  Hartree correction with respect to  the one of the  AA-p orbitals $\rm V_{H,AA-p}$, i.e. $\delta_{\rm H,a}=\rm V_{H,a}-V_{{\rm H},AA-p}$ with $\delta_{{\rm H},\rm AA-p}=0$ by definition. These Hartree corrections are found to be approximately linear in doping for both electron and holes. 
\begin{equation}
\delta_{{\rm H},a}=\delta^0_{{\rm H},a} +  V^*_{{\rm H},a} n
\label{eq:hartreefit}
\end{equation}
with $n$ the total number of electrons added or subtracted with respect to the CNP and $\delta^0_{{\rm H},a}$ a small correction found in undoped TBG. Fig.~\ref{fig:hartree} (c) and (d) show the values of $\delta^0_{{\rm H},a}$  
and $V^*_{{\rm H},a}$  found for $\epsilon=12$. The fittings have been performed for $n$ between -3 to 3 electrons. Clear deviations from linearity are found beyond this doping or for angles below $0.90^\circ$. As expected from the
sign of the $\Gamma-K$ shift, the proportionality factors $V^*_{{\rm H},a}$ are negative.

$V^*_{{\rm H},a}$ and, consequently,  $\delta_{{\rm H},a}$ depend  strongly on 
 the unknown value of the dielectric constant  $\epsilon$, see the Appendix. In atomistic calculations\cite{RademakerPRB2019, GoodwinArXiv2020} the relative shift of the states at the flat bands at $\Gamma$ and K is $\sim$ 6 meV 
 per electron or hole added to a free standing TBG. Those calculations are free from the uncertainty in the value of $\epsilon$. A suitable value of $\epsilon$ to describe the interactions in our model should reproduce the magnitude of the shift 
 between $\Gamma$ and K found in atomistic calculation, and controlled here by  $V^*_{{\rm H},a}$. As seen in Fig.~\ref{fig:hartree} the values obtained for $\epsilon=12$ are consistent with these findings.

Similar to the results found in Ref.~[\onlinecite{GoodwinArXiv2020}] there is a small but finite correction at the CNP. In our case this correction,  plotted in Fig.~\ref{fig:hartree} (d), is positive, i.e. it shifts the states of the flatbands upwards at $\Gamma$ with 
respect to the Dirac points, while in Ref.~\onlinecite{GoodwinArXiv2020} the shift at the CNP has an opposite sign. At present we do not know whether this difference is associated to the relaxation of the TBG considered here or 
to differences in the  Hartree reference which is subtracted in Eq.~(\ref{Eq:hartreepot}).

\section{Discussion and conclusions}
\label{sec:discussion}
The search for an interacting model based on effective moir\'e orbitals has been pursued since the discovery of the insulating and the 
superconducting states in TBG. The large number of atoms in the unit cell makes the analysis based on atomistic calculations complex and 
it is not obvious how to address Mott physics within calculations based on the continuum model, as the former involves the localization of electrons in real 
space and the suppression of quasiparticles in $k$-space. Multi-orbital effective models are specially suitable to address Mott correlations but also symmetry
breaking states, as the energy and length scales can be easily identified. However the analysis of the correlations was hampered by the complications of the 
early proposed two orbital model, such as the topological obstruction, the spinner function shape of the wavefunction, the slow decay of the interactions with distance 
and the sizable assisted hopping terms. 

In this work we have shown that the 8-orbital model, free from the topological obstructions by construction, features interactions more conventional than the 
ones of the two orbital one and can serve as the long awaited  interacting multi-orbital model for TBG.  Although this model includes a large number of 
orbitals, 16 when the two valleys are considered, four of them, the AA-p, deserve special attention as they contribute to the most part of the 
spectral weight of the flat bands and doped electrons will primarily fill them. We have shown that the onsite interaction between these AA-p orbitals $\sim 40$~meV is the largest 
interaction scale  in the model making them prone to Mott localization. This interaction is determined by the size of the AA region where the 
orbitals are centered. Contrary to common belief,  the size of the AA region, and not the one of the moir\'e unit cell, 
 controls the distance at which a gate has to be placed to screen this onsite interaction.  

The assisted hopping and exchange terms involving the AA-p orbitals are found to be small. Therefore the correlated states of TBG are expected to be governed by 
the density-density interactions. AA-p orbitals interact, not only among themselves, but also with the other orbitals included the model. These other orbitals are less correlated for twist angles 
$\theta \sim 1^\circ$ than the AA-p due to the larger bandwidth and smaller intraorbital interactions. The inter-orbital interactions decay as the inverse 
of the distance between their centers in the absence of gates and reaches $\sim 20$ meV for the interactions between the AA-p and AA-s centered at the same site. 

As a first approach to the correlations in TBG, we have treated the interactions at the Hartree level. The band deformation found in other approaches with doping is 
reproduced here and reveals the presence of orbitals different to AA-p at the $\Gamma$ point. We note that the magnitude of this effect depends not only on the interaction
between the AA-p orbitals but also on the interorbital interactions showing that on spite of the predominant role of the former, the interactions involving the other orbitals cannot be completely neglected. 
In summary our work paves the road for the study of correlations in TBG within the framework of multi-orbital models.  

We thank Zachary Goodwin for useful conversations. Funding from Ministerio de Ciencia, Innovaci\'on y Universidades (Spain) via grant  PGC2018-097018-B-I00 is gratefully acknowledged. 
        \renewcommand{\thesection}{APPENDIX}%
        \setcounter{table}{0} 
        \renewcommand{\thetable}{A\arabic{table}}%
        \setcounter{figure}{0}
        \renewcommand{\thefigure}{A\arabic{figure}}%
          \setcounter{equation}{0}
        \renewcommand{\theequation}{A\arabic{equation}}%
\section{Wave functions and Interactions} 

\subsection{Wannier functions}
For the calculations of the main text we start from the Wannier functions  $\Phi_{\alpha \tau l}({\bf r})$ corresponding to  the orbital $\alpha=1,...8$ of valley 
$\tau$ centered at a symmetry point of the moir\'e unit cell $l$. This Wannier function can be projected onto the four sublattices $X$, namely $A,B$ in layer $1,2$, 
and written in terms of an envelope function  $ \phi^{X}_{\alpha \tau l} ({\bf r}) $, with characteristic length scales of the order of 
the moir\'e lattice constant $\lambda$, and a Bloch factor $ \eta^{X}_{\tau} ({\bf r}) $ which oscillates at the atomic scale. 
\begin{equation}
\Phi_{\alpha \tau l}({\bf r})=\sum_X  \psi^{X}_{\alpha \tau l}({\bf r})= \sum_X  \phi^{X}_{\alpha \tau l} ({\bf r}) \eta^{X}_{\tau} ({\bf r}) 
\end{equation}
$ \phi^{X}_{\alpha \tau l} ({\bf r}) $ and $ \eta^{X}_{\tau} ({\bf r}) $ are complex functions and the Wannier functions of the two valleys 
$\Phi_{\alpha \tau l}({\bf r})$ are related by complex conjugation. The envelope functions $ \phi^{X}_{\alpha \tau l} ({\bf r}) $ were computed in Ref.~[\onlinecite{CarrPRR2019-2, kaxiraskp, kaxiras-github}] and have a different projection at each of the four sublattices $X$   while $ \eta^{X}_{\tau} ({\bf r}) $ 
is sublattice and valley dependent as~\cite{hauleArXiv2019}
\begin{equation}
\eta^{X}_{\tau} ({\bf r})=\frac{1}{\sqrt{3}}\sum_{j=3} e^{\pm i{\bf K}_j({\bf r} - {\bf R}_X)} 
\end{equation}
Positive (negative) sign applies to valley $+$ ($-$), the sum in $j$ refers to the three $K$ vectors of each, top or bottom, graphene layer and $R_X$ is the position of the carbon atom corresponding at each sublattice $X$ in the  graphene unit cell.  

\subsection{Interactions}
 In this work we have calculated the interactions between the Wannier functions
\begin{eqnarray}
H_{int}=\frac{1}{2} \sum_{\gamma_i,l_i,\sigma,\sigma'}U^{\gamma_a,l_a; \gamma_b,l_b}_{\gamma_d,l_d; \gamma_c,l_c} \left ( 1- \delta_{ad}\delta_{bc}\delta_{ab}\delta_{\sigma \sigma'}\right ) \times \nonumber  \\
d^\dagger_{\gamma_a l_a \sigma}d^\dagger_{\gamma_b l_b \sigma'} d_{\gamma_c l_c \sigma'} d_{\gamma_d l_d\sigma}
\end{eqnarray}
with $\gamma$ an index which includes the orbital and valley indices $\alpha$ and $\tau$, $\gamma_i=1,...,16$ and $l_i$ running to all the moir\'e unit cells. In the factor including the delta function we have used 
an abbreviated notation to avoid including the term with all indices equal, not allowed by Pauli exclusion principle. The interactions can be expressed as 
\begin{eqnarray}
U^{\gamma_a,l_a; \gamma_b,l_b}_{\gamma_d,l_d; \gamma_c,l_c} =\sum_{X X'} \int dr dr' V(|\bf{r}-\bf{r}'|) \times \nonumber \\
\psi^{X*}_{\gamma_a l_a}({\bf r}) \psi^{X'*}_{\gamma_b l_b}({\bf r}') \psi^{X'}_{\gamma_c l_c}({\bf r}') \psi^{X}_{\gamma_d l_d}({\bf r})
\label{apeq:formulaU}
\end{eqnarray}
where we neglect any overlap of the carbon $p_z$-orbitals in different sites.  Here  $V(|\bf{r}-\bf{r}'|)$, the interaction between the electrons in the carbon atoms, is given by
\begin{equation}
V(|{\bf r}-{\bf r}'|)=\frac{e^2}{4\pi \epsilon_0 \epsilon} \frac{1}{(|\bf{r}-\bf{r}'|)}
\label{eq:potcoulomb}
\end{equation} 
in the absence of gates, with $|\bf{r}-\bf{r}'|$ the distance between the electrons in the $2D$ plane. 
The magnitude of the interactions depends on $\epsilon$ as $1/\epsilon$. $\epsilon$ accounts for the internal screening of the TBG due to the electrons in the high energy bands not included in the 8-orbital model and for an extra external screening if the TBG is 
encapsulated in h-BN. The value of $\epsilon$ is a priori not known,  however, as discussed above, the comparison of the Hartree correction in our model and in atomistic calculations allows us to estimate $\epsilon \sim 12$ in the absence of h-BN.  

The screening by metallic gates is accounted for by means of the images charge method.
$V(|\bf{r}-\bf{r}'|)$ becomes
\begin{equation}
V_{sg}(|{\bf r}-{\bf r}'|)=\frac{e^2}{4\pi \epsilon_0 \epsilon} \left(\frac{1}{(|{\bf r}-{\bf r}'|)}-\frac{1}{\sqrt{(|{\bf r}-{\bf r}'|)^2+(2d)^2}}\right)
\end{equation}
in proximity to a single gate at distance $d$ and 
\begin{eqnarray}
V_{dg}(|{\bf r}-{\bf r}'|)& = & \frac{e^2}{4\pi \epsilon_0 \epsilon} \sum_{n=-\infty}^{\infty} \frac{(-1)^n}{\sqrt{|{\bf r}-{\bf r}'|^2 +(2dn)^2}}  \nonumber \\ 
             & \sim & \frac{e^2}{4\pi \epsilon_0 \epsilon} \frac{2 \sqrt{2}e^{-\pi |{\bf r}-{\bf r}'|/2d}}{\sqrt{2 d |{\bf r}-{\bf r}'|}}
\end{eqnarray}
when the TBG is placed at equal distance $d$ of two metallic gates \cite{ThrockmortonPRB2012}. The last approximated expression is valid only for distances larger than $d/2$. 
In all cases we regularize the interaction at ${\bf r}={\bf r}'$ to $28.7/\epsilon$ eV, equivalent in the absence of gates to the interaction between two electrons at distance $0.35 a_0$ with $a_0$ the distance between two carbon atoms.

Density-density interactions, discussed in the main text, refer to interactions in Eq.~($\ref{apeq:formulaU}$) satisfying $\gamma_a=\gamma_d$, $l_a=l_d$ and $\gamma_b=\gamma_c$, $l_b=l_c$.  Density-density interactions are intra-orbital if  $\gamma_a=\gamma_b$ and intra-unit cell if $l_a=l_b$. Onsite interactions are  both intra-orbital and intra-unit cell. The contribution of density-density interactions to the Hamiltonian can be written as 
\begin{equation}
H_{dens}=\frac{1}{2} \sum_{\gamma_a,\gamma_b,\sigma_a,\sigma_b}U^{\gamma_a,l_a}_{\gamma_b,l_b} (1-\delta_{ab}\delta_{\sigma_a \sigma_b}) n_{\gamma_al_a \sigma_a} n_{\gamma_b l_b \sigma_b}
\label{apeq:Hdens}
\end{equation}
with $n_{\gamma_il_i\sigma_i}=d^\dagger_{\gamma_i l_i \sigma_i} d_{\gamma_i l_i \sigma_i}$ the electronic density  of the orbital $\gamma_i$, centered at unit cell $l_i$ and with spin $\sigma_i$ and
\begin{equation}
U^{\gamma_a,l_a}_{\gamma_b,l_b}=U^{\gamma_a,l_a; \gamma_b,l_b}_{\gamma_a,l_a; \gamma_b,l_b}
\end{equation} 
In the main text we use the simplified notation $U^j_{a,b}$ to refer to the density-density interactions between orbitals $a$ and $b$, with this index including valley, centered at unit-cells whose centers are separated by $j$ times the moir\'e  lattice constant $\lambda$. In particular $U^0_{a,b}$ is the onsite interaction and $U^1_{a,b}$ the interaction between orbitals in unit-cells which are first-neighbors. Some care is needed, as the interaction between the centers at the hexagonal or kagome lattice of two different unit cell, may depend not only on the distance between the centers of the two unit cells, but also on the relative orientation of these unit-cells. 

\begin{figure}
\leavevmode
\includegraphics[clip,width=0.45\textwidth]{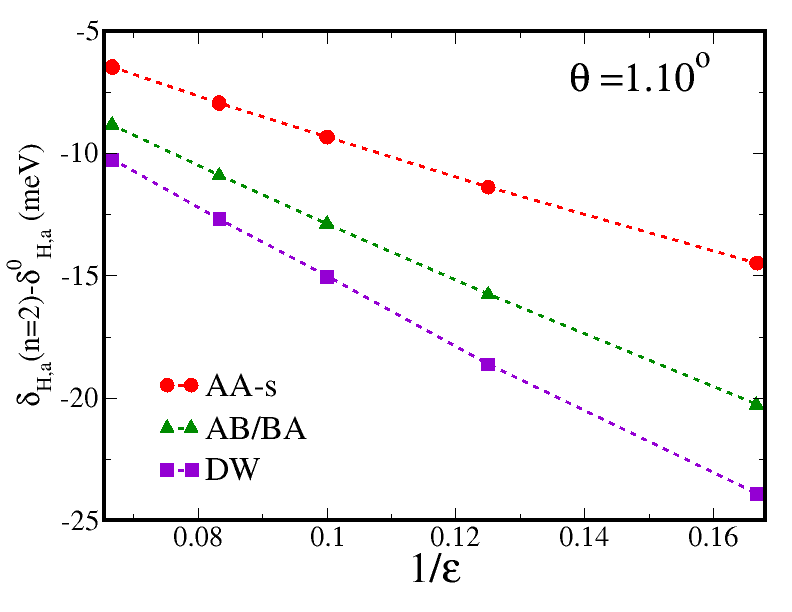}
\caption{Orbital dependent hartree corrections with respect to CNP for a $\theta=1.10^\circ$ TBG doped with two electrons versus the inverse of the dielectric constant. } 
\label{fig:hartreeappendix}
\end{figure}

In the exchange interactions $\gamma_a=\gamma_c$, $l_a=l_c$, $\gamma_b=\gamma_d$, $l_b=l_d$ and $\sigma_a=\sigma_b$.  
\begin{equation}
J^{\gamma_a,l_a}_{\gamma_b,l_b}=U^{\gamma_a,l_a; \gamma_b,l_b}_{\gamma_a,l_a; \gamma_b,l_b}
\end{equation}
\\
In the text we use a simplified notation $J^j_{ab}$ analogous to the one defined for the density-density interactions.
Exchange interactions involving electrons in Wannier functions centered at the same site are here denoted as Hund's coupling.
If $\gamma_a$ and $\gamma_b$ are in different valleys, the product of the Bloch factors oscillates as $e^{i({\bf K}-{\bf K})'{\bf r}}$, with ${\bf K},{\bf K}' $  the two inequivalent graphene $\bf K$ points. This product oscillates at the atomic
scale integrating to a small value. 

Finally, we  have also studied the assisted hopping terms with $\gamma_a=\gamma_d$, $l_a=l_d$ and $\gamma_b \neq \gamma_c$ and/or $l_b \neq l_c$. 
\subsection{Hartree approximation}
To calculate the Wannier potentials in Eq.(\ref{Eq:hartreepot}) we use
\begin{equation}
V_{H a,\sigma_a} =\sum_{\gamma_b, l_b,\sigma_b} U^{\gamma_a,l_a}_{\gamma_b,l_b} (1-\delta_{ab}\delta_{\sigma_a \sigma_b})(\left < n_{\gamma_bl_b\sigma_b}\right> - \bar n)
\label{eq:hartreepotap}
\end{equation}
with $\left < n_{\gamma_bl_b\sigma_b}\right>$ calculated self-consistently, updating $V_{H a,\sigma_a}$ at each self-consistency step.
In the sum we include the closest 54 nearest neighbors unit cells $l_b$. We  calculate $U^{\gamma_a,l_a}_{\gamma_b,l_b}$ for $l_b=l_a$ and for $l_b$ corresponding to the six closest unit cells using the Wannier functions, and approximate $U^{\gamma_a,l_a}_{\gamma_b,l_b}$ as given by a point charge placed at the corresponding Wannier funcion centers for the other unit cells $l_b$. The interaction potential (\ref{eq:potcoulomb}) is used in the calculation  and the  interactions $U^{\gamma_a,l_a}_{\gamma_b,l_b}$ which enter in Eq.(\ref{eq:hartreepotap}) are proportional to $1/\epsilon$. The orbital dependent Hartree corrections $\delta_{H,a}$ inherit approximately this dependence as shown  in Fig.~(\ref{fig:hartreeappendix}).

\bibliography{8band}
\end{document}